\documentclass[preprint,12pt,authoryear]{elsarticle}

\usepackage{amssymb}
\usepackage{amsmath}
\usepackage{booktabs}

\usepackage{eurosym}

\usepackage{float}
\usepackage{color,soul}
\usepackage{graphicx}

\usepackage{hyperref}
\hypersetup{breaklinks=true}

\journal{Journal of Transport Geography}

\begin{document}

\begin{frontmatter}


\title{Mapping Travel Experience in Public Transport: Real-Time Evidence and Spatial Analysis in Hamburg}

\author[DLR]{Esther Bosch\corref{cor1}}
\ead{Esther.Bosch@dlr.de}
\ead[url]{https://orcid.org/0000-0002-6525-2650}

\author[DLR]{Michael Scholz}
\ead[url]{https://orcid.org/0000-0002-1052-6503}

\author[DLR]{Anke Sauerländer-Biebl}
\ead[url]{https://orcid.org/0000-0002-1052-6503}

\author[DLR]{Klas Ihme}
\ead[url]{https://orcid.org/0000-0002-7911-3512}

\cortext[cor1]{Corresponding author}

\address[DLR]{German Aerospace Center (DLR), Institute of Transportation Systems,\\
Lilienthalplatz 7, 38108 Braunschweig, Germany}

\begin{abstract}
Shifting travel from private cars to public transport is critical for meeting climate and related mobility goals, yet passengers will only choose transit if it offers a consistently positive experience. Previous studies of passenger satisfaction have largely relied on retrospective surveys, which overlook the dynamic and spatially differentiated nature of travel experience. This paper introduces a novel combination of real-time experience sampling and spatial hot spot analysis to capture and map where public transport users report consistently positive or negative experiences.

Data were collected from 239 participants in Hamburg between March and September 2025. Using a smartphone application, travelers reported their momentary journey experience every five minutes during everyday trips, yielding over 21,000 in-situ evaluations. These geo-referenced data were analyzed with the Getis-Ord $Gi^{*}$ statistic to detect significant clusters of positive and negative travel experience. The analysis identified distinct hot and cold spots of travel experience across the network. Cold spots were shaped by heterogeneous problems, ranging from predominantly delay-dominated to overcrowding or socially stressful locations. In contrast, hot spots emerged through different pathways, including comfort-oriented, time-efficient or context-driven environments. 

The findings highlight three contributions. First, cold spots are not uniform but reflect specific local constellations of problems, requiring targeted interventions. Second, hot spots illustrate multiple success models that can serve as benchmarks for replication. Third, this study demonstrates the value of combining dynamic high-resolution sampling with spatial statistics to guide more effective and place-specific improvements in public transport.

\end{abstract}

\begin{graphicalabstract}

\includegraphics[width=1\linewidth]{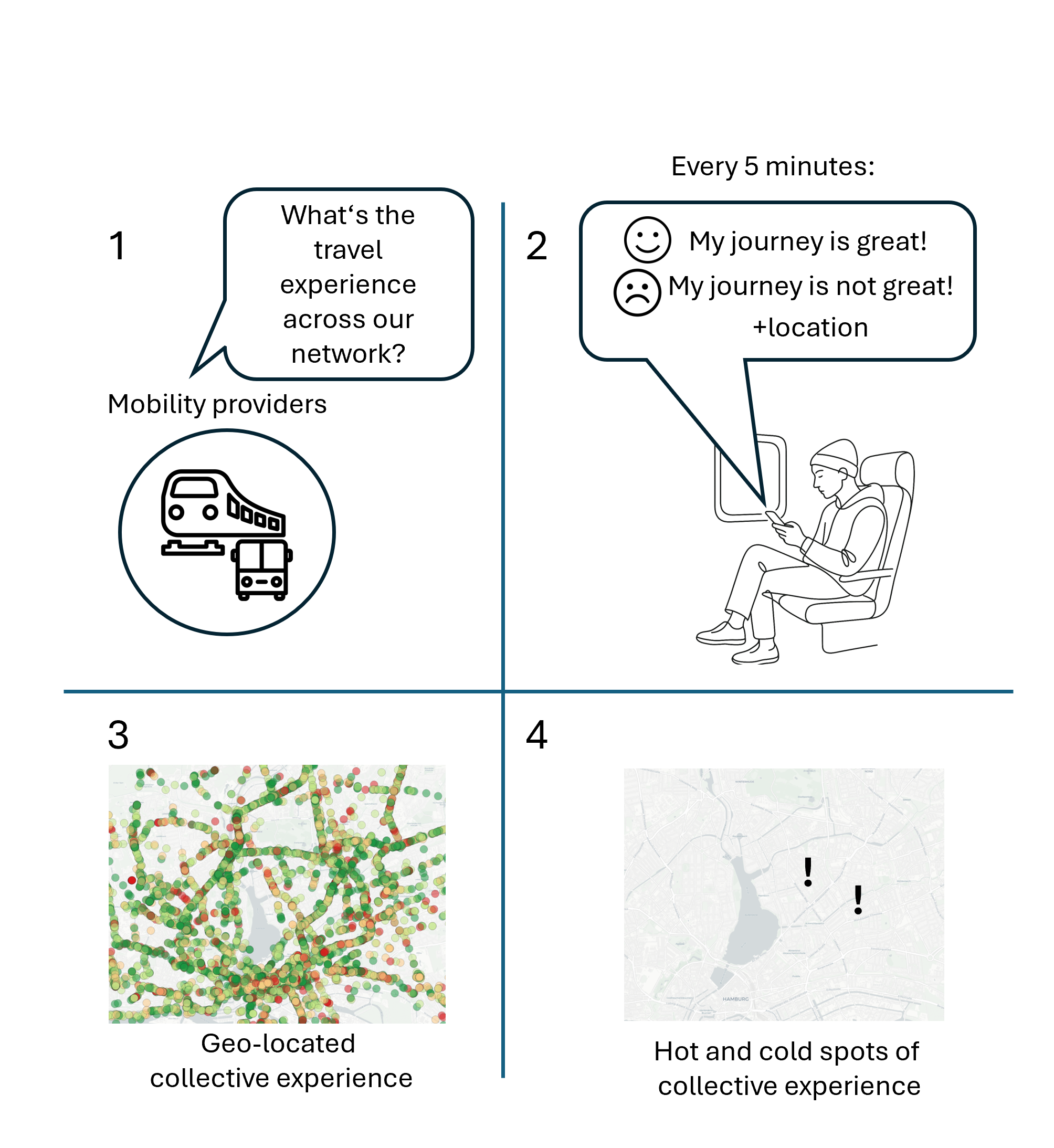}

\end{graphicalabstract}

\begin{highlights}
\item Real-time experience sampling collected over 21,000 in-situ travel evaluations
\item Getis-Ord $Gi^{*}$ detects spatial hot and cold spots of travel experience
\item Cold spots differ: delay-dominant vs. overcrowding and social stress locations
\item Hot spots emerge through comfort, time efficiency, or pleasant environment
\item Negative events outweigh positives, highlighting priorities for targeted action
\end{highlights}


\end{frontmatter}



\section{Introduction}

Transportation accounts for 23\% of global energy-related CO$_{2}$ emissions \citep{IPCC_AR6_WG3_Ch10_2022} and therefore, a shift towards sustainable transportation is essential to meet climate neutrality goals \citep{EU2021climatelaw}. Public transportation produces far fewer emissions per passenger than cars \citep{noussan2022carbon} and can therefore significantly reduce the transport sector’s carbon footprint. However, people will only choose public transport if it offers a convenient and attractive travel experience \citep{de2022attitude}. This raises the need to improve public transport so that it becomes a mode of choice for more travelers, thereby advancing both sustainability and mobility goals.

Extensive research has examined what factors or events during a journey influence travelers’ satisfaction with public transport \citep{van2018influences}. Studies have identified service reliability and frequency of departures as particularly important for making transit attractive \citep{goransson2023factors, redman2013quality}. Other key determinants of satisfaction include accessibility, comfort, staff behavior, and perceived safety \citep{van2018influences}. These service attributes collectively shape the travel experience that results in an overall travel satisfaction \citep{lim2024effects}. High satisfaction not only improves current riders’ well being \citep{chatterjee2020commuting} but can also encourage continued ridership \citep{carvalho2022loyalty} and modal shift from cars \citep{abou2012travel}, especially when transit services meet or exceed user expectations. This perspective aligns with Kahneman’s concept of experienced utility, which emphasizes the direct measurement of individuals’ moment-to-moment affective states during an experience, rather than relying solely on remembered or decision-based evaluations \citep{kahneman1997back}. Additionally, \citet{cornet2022worthwhile} introduce the concept of worthwhile travel time, emphasizing that travel can itself be pleasant, meaningful, and valuable rather than merely a disutility to be minimized. These findings, in summary, highlight the relevance, but also potentials, of improving travel experience in order to motivate a long-term modal shift towards public transport.

Given the multitude of factors affecting passenger satisfaction, transit operators face the challenge of prioritization. Agencies typically operate under constrained budgets and personnel, so they cannot address every possible issue at once. It is therefore crucial to identify which improvements will yield the largest gains in rider satisfaction and attract the most new users. In practice, this means pinpointing the critical incidents \citep{allen2020effect, friman2001frequency} and pain points in the travel experience \citep{lim2024effects}, and focusing investment there. Past studies have provided general insights into what matters to riders \citep{allen2020effect, goransson2023factors, redman2013quality, van2018influences}, but finding out where and what interventions would be most effective remains difficult. Transit experiences are not uniform across a city; they can vary widely by route, location, and time of day. For example, one neighborhood’s bus stop might routinely suffer overcrowding and delays, while another area’s service runs smoothly. Understanding the geography of travel experience is thus an important next step for maximizing the impact of improvements. 

To achieve a geographically differentiated understanding of travel experience, it is essential to collect experiential data together with the precise location of each reported moment. Recent advances in data-driven mobility analytics provide powerful tools to capture and understand how infrastructure and service design shape travel behavior and experience in real-world contexts. Earlier work in the field of affective geovisualization includes Nold's Emotional Cartography, in which bodily and locational traces of emotion are mapped to understand the affective dimension of everyday urban mobility and experience \citep{nold2009emotional}. Similarly, the Urban Emotions framework \citep{resch2014urban} demonstrated how emotional responses can be spatially captured from human and technical sensors to inform urban planning. These approaches illustrate how spatial analytics can uncover collective affective patterns in urban environments. \citet{metz2024interactive} present an interactive visual analytics framework that combines open data sources with population-based mobility simulations to assess accessibility and network quality at the housing level. These approaches show how computational and visualization methods can complement psychological and experiential perspectives by revealing where transport networks succeed or fail to meet diverse user needs.


Using such geospatial information on emotion and experience enables to calculate areas of collective high and low experience. For example, \citet{cardone2022gis} showed that GIS-based hot and cold spot detection can be extended to emotional data from social media. Among the available spatial clustering techniques \citep{anselin1995local, mcqueen1967some, devroye2001combinatorial}, the Getis–Ord Gi* statistic has the advantage that it directly identifies localized clusters of high or low values based on the intensity of neighboring points, allowing a precise delineation of areas where travelers collectively report unusually positive or negative experiences without the need to predefine cluster shapes or boundaries. 

Combining these previous findings and methods, our study employs an in-situ experience sampling method (ESM) \citep{hektner2007experience} to collect real-time travel experience reports at frequent intervals during actual trips. In our experiment, 239 public transport users provided feedback on their travel experience every five minutes throughout their everyday journeys. On this data, we apply the Getis-Ord $Gi^*$ hot spot statistic to the geo-referenced experience reports. The Getis-Ord $Gi^*$ statistic detects where values are clustered spatially, in our case, identifying locations that register collectively high or collectively low experience ratings \citep{GetisOrd1992}. Using this technique, we uncovered distinct geographic clusters where travelers collectively report very positive or very negative trip experiences. This represents, to our knowledge, the first application of formal spatial statistics to a large-scale travel experience dataset in public transport.

Combining the spatial hot spot analysis with the rich ESM data on reported travel experience and trip events allows us to explore why certain areas are performing poorly or well. We can identify which specific experience-influencing events (for instance, delays, overcrowding, disruptive people, missed connections, etc.) occur frequently in the locations that passengers collectively rate with a negative travel experience. In effect, our results create a map of the transit system’s collective pain points and highlight what causes them. This offers valuable, actionable insights for mobility operators and planners. With our findings, transit agencies can see where travelers are most dissatisfied and what tends to be happening in those areas to trigger that dissatisfaction. Such information is crucial for guiding investments. For example, if a particular area is a cold spot due to chronic overcrowding events, resources can be directed to increase capacity or frequency there, rather than spreading resources too thinly system-wide.

In summary, this study makes both a substantive and a methodological contribution. Substantively, it provides new evidence on the factors driving travel experience at a very granular level and identifies targeted areas for improvement in the pursuit of more sustainable, traveler-friendly transport. Methodologically, it demonstrates the power of combining real-time experience sampling with spatial statistical analysis in transport geography, as suggested by \cite{Martino2010SenseableCity,bosch2025travel,bosch2023travel}. By this, we offer a novel approach for fellow researchers to explore collective travel satisfaction and pinpoint issues in other contexts. Our introduction of Getis-Ord hot spot analysis to travel experience research opens up a new avenue for understanding how the quality of transit service varies across space, and how that spatial perspective can inform more effective interventions.

\section{Methods}

\subsection{Procedure}
Data were collected in Hamburg, Germany, between March and September 2025. Hamburg is Germany’s second-largest city, with approximately 1.8 million inhabitants and a metropolitan region of around 5.5 million people. Located in the country’s central north on the river Elbe, Hamburg serves as a major economic and transport hub, combining a dense urban core with extensive suburban areas. The city has a comprehensive public transport network operated under the Hamburger Verkehrsverbund (HVV), including subway (U-Bahn), suburban rail (S-Bahn), buses, and ferries, providing highly interconnected coverage across the metropolitan region. 
\\
For the study, participants were instructed to record at least six everyday travel routes of a minimum duration of 15 minutes each within a period of three weeks from enrollment. 
Participants received a compensation of \EUR{70} for their participation if they successfully recorded six routes of at least 15 minutes. If fewer routes were recorded or questionnaires not always answered, the compensation was pro-rated according to the number of completed eligible routes and questionnaires. This was communicated to the participants before the start of the study.

\subsection{Participants}

239 participants were recruited via advertisements on in-vehicle screens in the city’s subway. In the selection of participants, efforts were made to mitigate potential recruitment biases by balancing the sample in terms of age and gender. The final sample consisted of 1 participant identifying as diverse, 117 as male, 116 as female, and 5 who preferred not to disclose their gender. The mean age was 34 years ($\pm$ 11). Among those who provided information on transport habits, 4 reported using public transport once per week, 78 several times per week, and 155 daily.  

Before the start of the study, participants were informed about the study procedures and data handling practices, and gave written consent in accordance with the General Data Protection Regulation (GDPR). The study was reviewed and approved by the institution’s ethical committee (reference no. 04/25).

\subsection{Data Collection}

After completing the informed consent form and a demographic questionnaire, participants were instructed to download the institute’s research smartphone app. 
Prior to each journey, participants selected their intended route within the app. In this way, data were collected both on the set of possible routes and the chosen route, including line numbers and departure times.  

Before the start of each trip, participants were asked to indicate whether they had any special requirements (e.g., accessibility needs, traveling with children, carrying luggage).  

During the trip, participants received a pop-up questionnaire every five minutes via the app. This instrument is referred to as the 'Experience Sampling Questionnaire' throughout this paper. The following questions were answered on sliders under the heading 'Evaluation of journey' as translated from German:  

\begin{enumerate}
    \item with the current situations on my journey I am 1 (totally dissatisfied) to 5 (totally satisfied)
    \item the journey at the moment is 1 (miserable) to 5 (excellent)
    \item for achieving my goals, this journey is 1 (a hindrance) to 5 (favourable)
    \item at the moment my journey is going 1 (not smoothly) to 5 (smoothly)
    \item I currently find my journey 1 (unpleasant) to 5 (pleasant)
    \item compared to an ideal trip, my trip so far is 1 (the worst) to 5 (the best)
    \item the current situation on my journey is 1 (worse than expected before starting to travel) to 5 (better than expected before starting to travel)
    \item based on my previous experience with traveling, this journey is 1 (below average) to 5 (above average)
\end{enumerate}

The construct 'travel experience' was calculated per experience sampling questionnaire by taking the mean value of the responses to the above eight questions.

On the same screen, participants could set tick marks next to any events that occurred since the last questionnaire.  The choice of event categories was taken from previous research using experience sampling in public transport \citep{bosch2023happens, bosch2023travel}. A text above explained: 'Which factors currently influence your evaluation of the journey?'

\begin{itemize}
    \setlength{\itemsep}{0pt}
    \setlength{\parskip}{0pt}
    \item delay
    \item missed connection
    \item had to hurry
    \item disturbing people are around me (called 'disruptive people' throughout the paper)
    \item crowded vehicle (called 'overcrowded')
    \item driving behavior of the driver (called 'driving behavior')
    \item infrastructure (for example, no roof, not barrier-free, uneven street) (called 'infrastructure issues')
    \item not enough information (called 'missing information')
    \item positive social interaction or situation (called 'positive interaction')
    \item media time well spent (music, podcast, book, etc) (called 'time well spent')
    \item I reached my destination or connection or the means of transport is on time (called 'arrived on schedule')
    \item discomfort (bored, impatient, difficult thoughts, hungry, tired, cold, wet, etc) (called 'feeling unwell')
    \item means of transport is comfortable (cosy, good temperature, not too crowded, got a seat, etc) (called 'comfort')
    \item beautiful environment (called 'nice environment')
    \item other - with the option of a free text input
\end{itemize}

Each of these questionnaires was stored with a time stamp and a GNSS-based location.

At the end of each journey, the app automatically opened the post-trip questionnaire that asked several questions about the overall evaluation of the trip.

To analyze the free-text responses, three independent raters categorized all answers. Inter-rater reliability was assessed using Fleiss' Kappa across three raters and 17 categories, yielding a value of $\kappa = .73  (SD = 0.14)$, indicating substantial agreement \citep{landis1977measurement}, with per-category kappas ranging from $\kappa = .54$ (others) to $\kappa = .90$ (polite staff). In cases of disagreement, the raters discussed their assessments and reached a consensus through deliberation.

\subsection{Confirmatory Factor Analysis}
Confirmatory factor analysis was conducted in R using the \texttt{lavaan} package \citep{rosseel2017package}, with the WLSMV estimator to account for the ordinal nature of the indicator variables. Model fit was evaluated using standard criteria (CFI, TLI, SRMR, RMSEA), and reliability was assessed via McDonald's Omega Hierarchical ($\omega_H$), computed from standardized loadings and residual variances.

\subsection{Spatial analysis}

To identify spatial clusters of high and low travel experience, we applied the
Optimized Hot Spot Analysis tool in ArcGIS Pro \citep{esri_optimized_hotspot_2025}. This procedure is based on the Getis-Ord $Gi^{*}$ statistic, which detects where high or low values are spatially concentrated beyond what would be expected by random chance. A location is classified as a hot spot when it has a high value surrounded by other high values, and as a cold spot when it has a low value surrounded by other low values. Importantly, the method evaluates spatial clustering independently of how many participants were sampled at a given location, which makes it suitable for uneven data distributions. Statistical significance is adjusted for multiple testing and spatial dependence using the False Discovery Rate (FDR) correction method.

The Optimized Hot Spot Analysis determines the appropriate scale of analysis by first running the Incremental Spatial Autocorrelation tool. This tool applies the Global Moran’s I statistic to the data at a series of increasing distances, calculating a $z$-score that reflects the intensity of spatial clustering at each distance. Typically, clustering intensifies with distance until a peak is reached, after which the $z$-score declines. The peak distance indicates the scale at which spatial processes promoting clustering are most pronounced, and this distance is then used as the fixed distance band in the $Gi^{*}$ analysis.
Locational outliers are excluded when computing the initial and incremental
distances. As a result, the optimized analysis automatically identifies the
spatial scale at which clustering is strongest, ensuring that the detection of
hot and cold spots reflects the most relevant geographic patterns in the data. 

The hot and cold spots were numbered from 1 to 12 by manually drawing polygon features that tightly enclosed the spatial extent of each collective hot and cold spot.

\section{Results}

\begin{figure}[H]  
\centering
\includegraphics[width=1\linewidth]{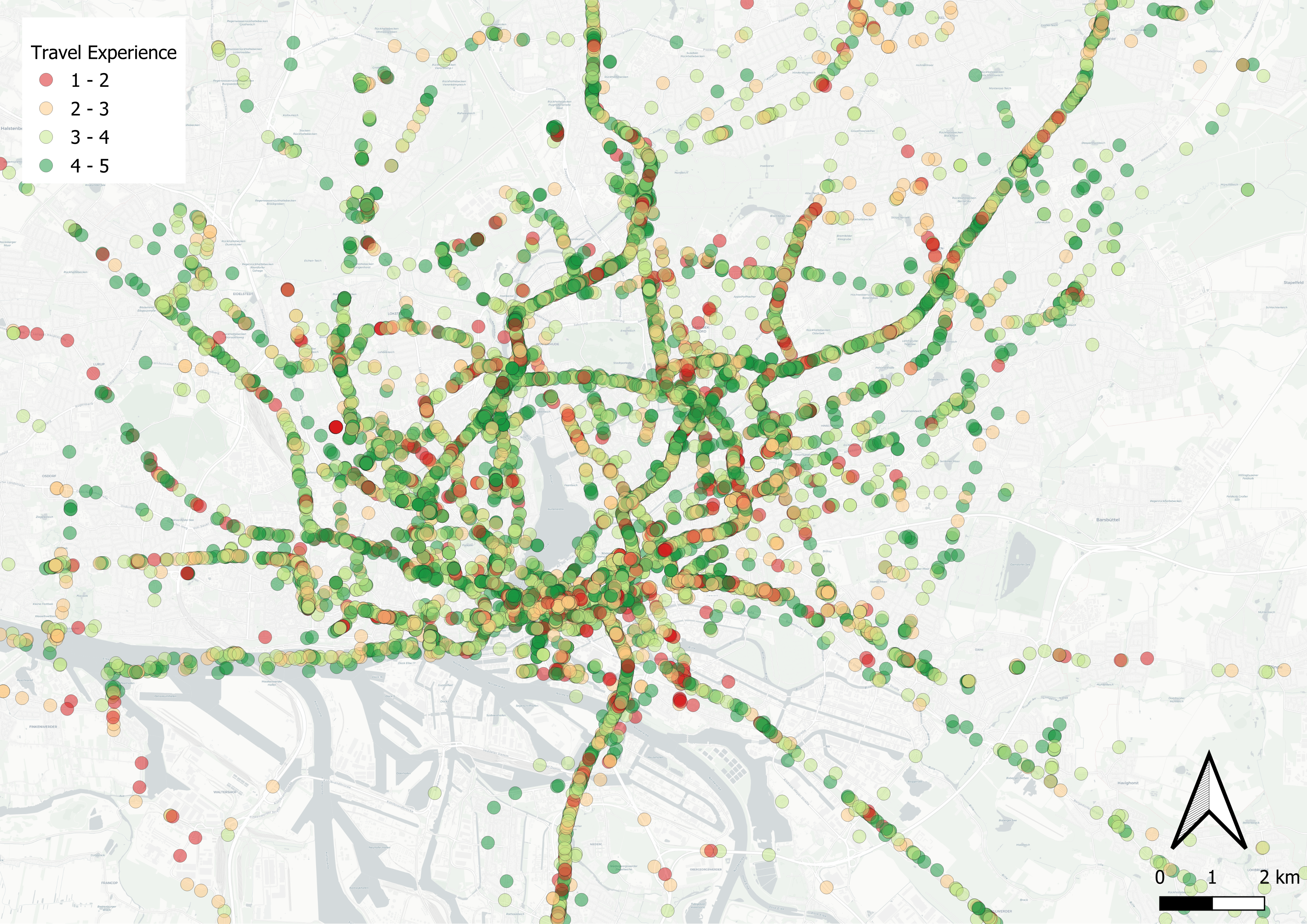}
\caption[Raw data overview]{\label{fig:exp}Raw data overview. Base map \copyright{} OpenStreetMap community (ODbL).}
\end{figure}

On average, participants reported 13.49 $\pm$ 34.01 drives and answered 73.06 $\pm$ 74.97 questionnaires during their trips overall. Over all participants, 21{,}411 questionnaires during trips were answered. 
Figure \ref{fig:exp} gives an exemplary insight of the data collected in Hamburg's inner city. 

\subsection{Experience Sampling Questionnaire}
Descriptive analyses of the 8-item travel experience questionnaire revealed item means ranging from 2.47 to 4.89 ($SD = 0.07$ -- $1.23$). 

Although the nested data structure would theoretically call for a three-level confirmatory factor analysis \citep{Bohmann_under_preparation}, we employed a standard CFA based on each participant's first measurement from their first trip ($n = 301$). Robustness checks using either the last or a randomly selected measurement produced comparable results.

A CFA with ordinal indicators (WLSMV estimator) indicated good model fit ($\chi^{2}(18) = 36.99$, $p = .005$; CFI $= .999$; TLI $= .999$; SRMR $= .032$; RMSEA $= .059$, 90\% CI $[.031, .086]$; robust CFI $= .939$; robust TLI $= .906$; robust RMSEA $= .179$). The final model specifies momentary travel experience as a higher-order factor, with items 1, 2, and 5 serving as its direct indicators. Comparative travel experience (items 6--8) and goal-oriented travel experience (items 3--4) were modeled as first-order facets loading onto this higher-order factor. Standardized loadings were uniformly high (.85--.94), and the hierarchical factor accounted for the majority of variance. One- and two-factor alternatives---including a two-factor model corresponding to our initial hypothesis---fit the data less well, confirming that the hierarchical specification offers the most parsimonious and accurate representation of the dimensionality of travel experience. Reliability was assessed via McDonald's Omega Hierarchical computed from standardized loadings and residual variances, yielding $\omega_{H} = .83$.

\subsection{Getis-Ord Gi*}

\begin{figure}[H]  
\centering
\includegraphics[width=1\linewidth]{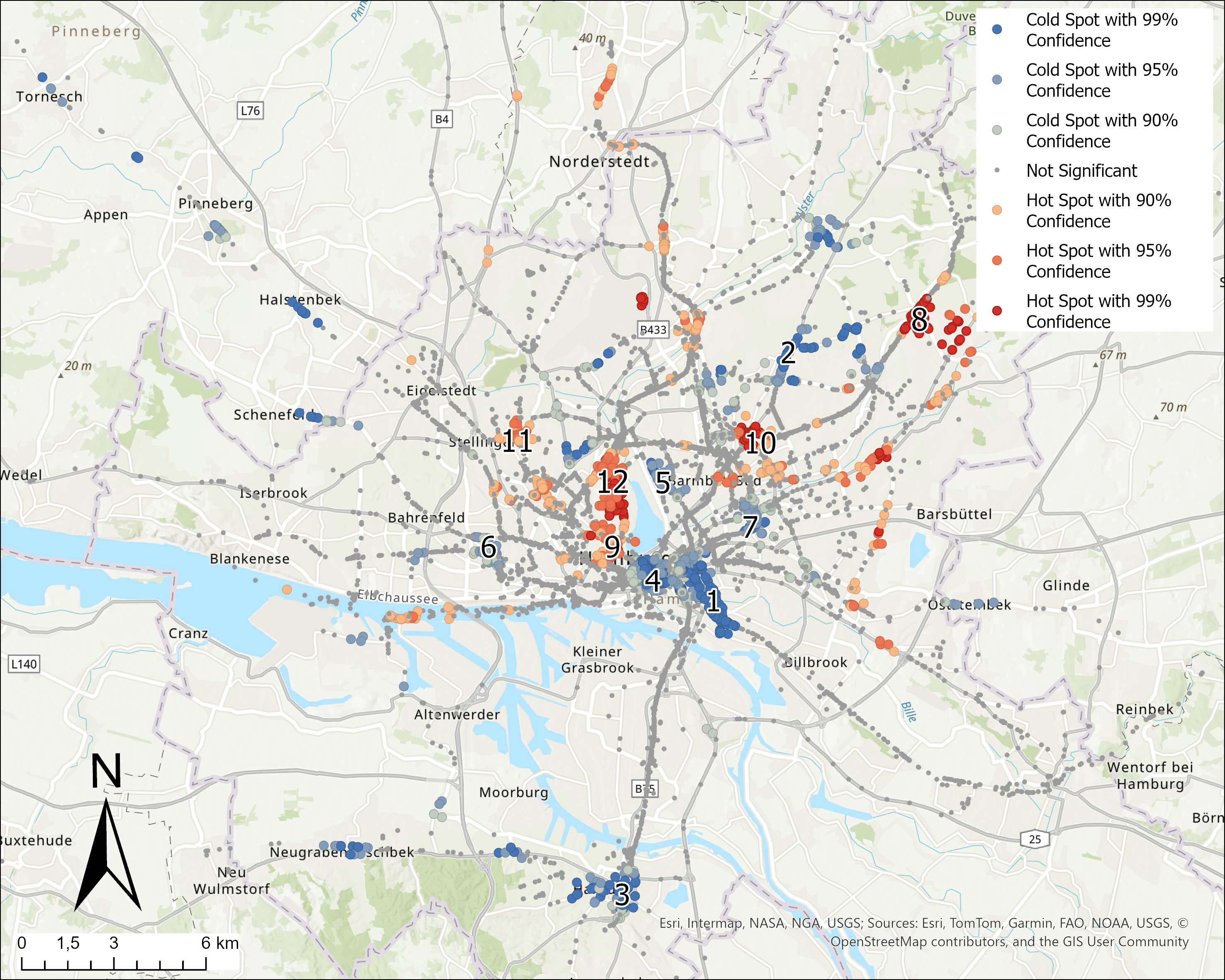}
\caption[Hot and cold spots calculated by Optimized Getis-Ord $Gi^{*}$ statistic]{\label{fig:hotspots}Hot and cold spots calculated by Optimized Getis-Ord $Gi^{*}$ statistic. Numbers 1--12 label individual hot and cold spot areas for reference throughout the paper; they do not imply ranking or ordering.}
\end{figure}

\begin{figure}[H]  
\centering
\includegraphics[width=1\linewidth]{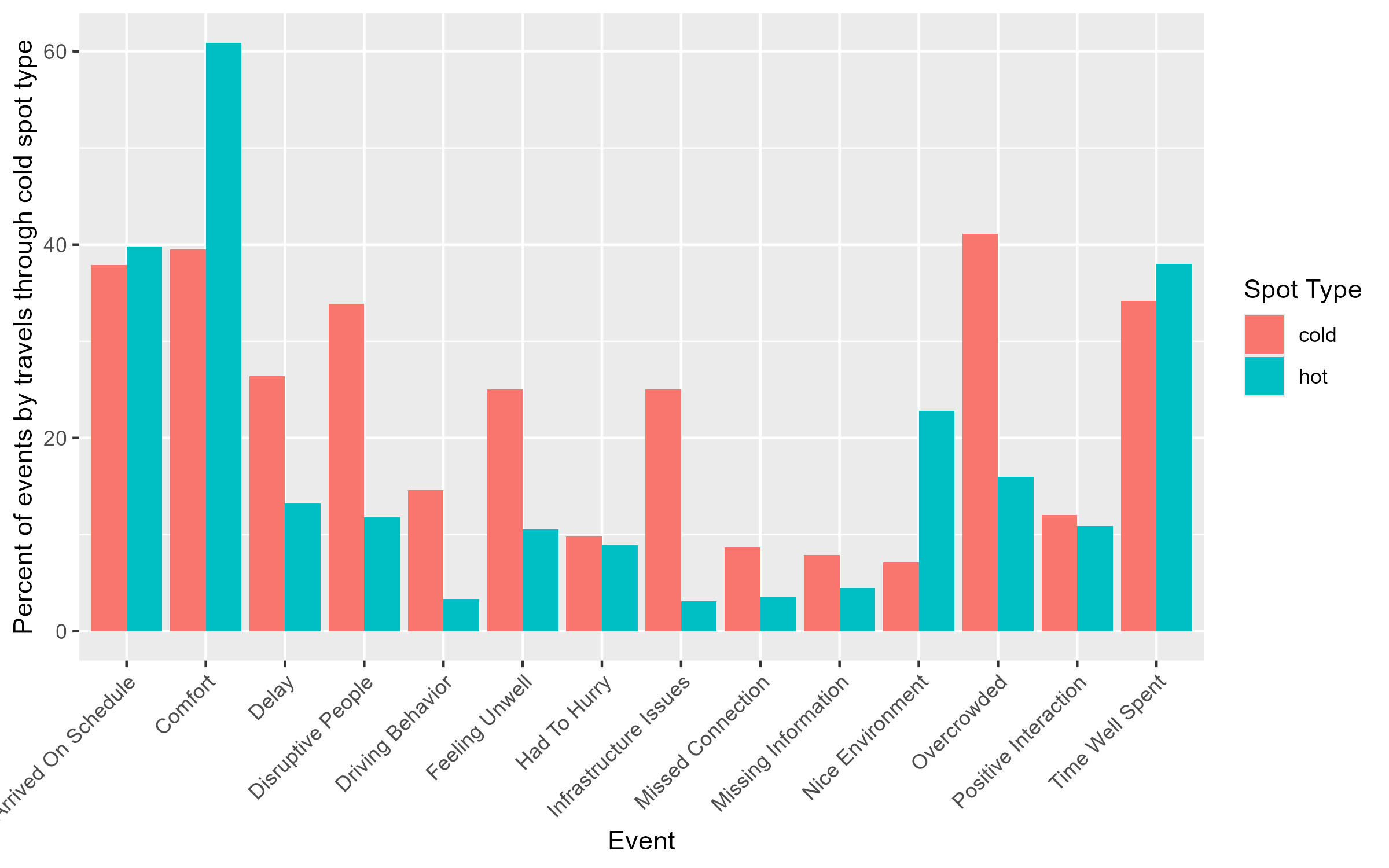}
\caption{\label{fig:eventsperspottype}Counts of reported events during experience sampling per hot or cold spot.}
\end{figure}

\begin{figure}[H]  
\centering
\includegraphics[width=1\linewidth]{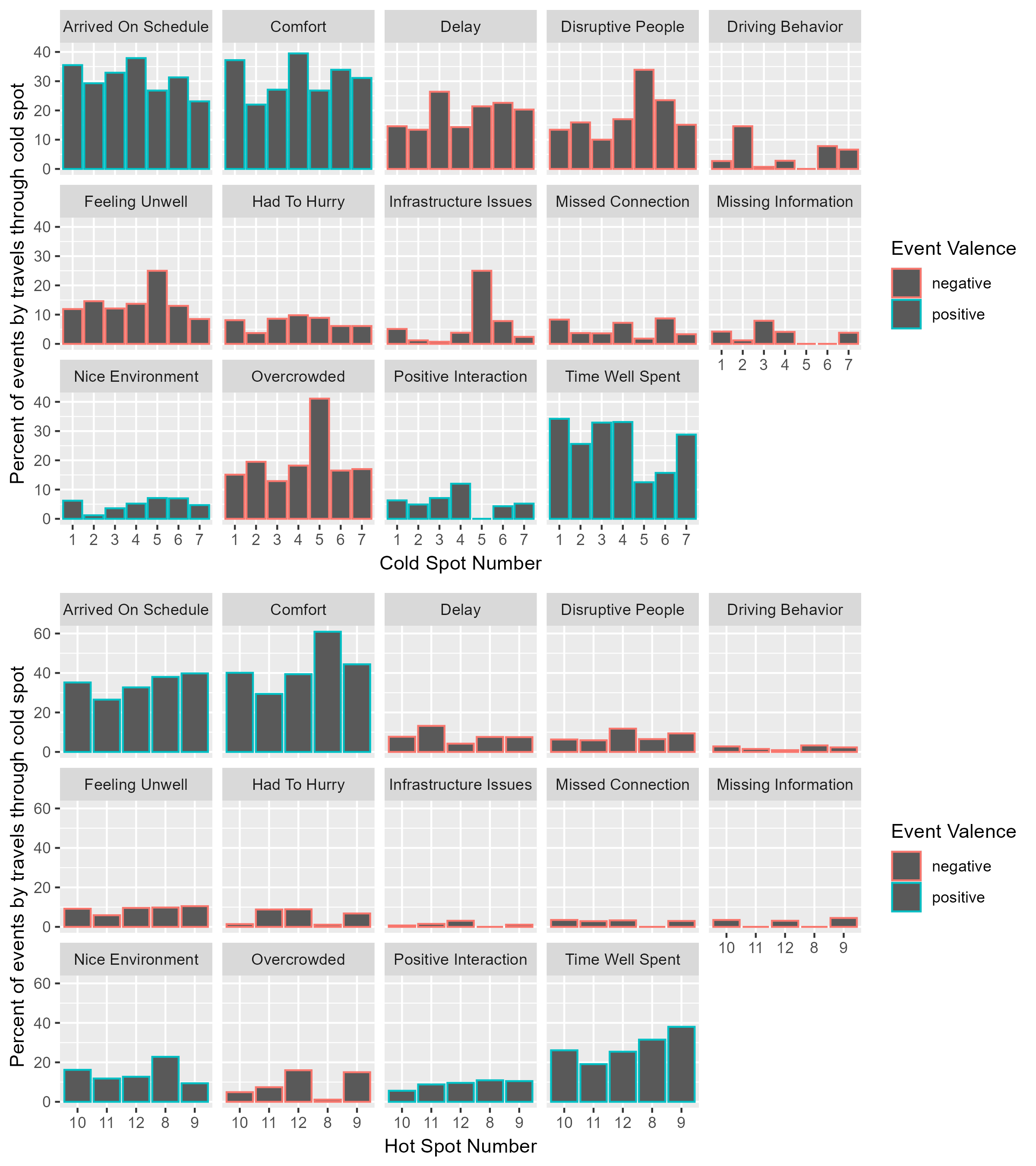}
\caption{\label{fig:hotspotsevents}Counts of reported events during experience sampling per hot or cold spot.}
\end{figure}

Figure~\ref{fig:hotspots} reveals spatially distinct clusters of momentary travel experience. Using a fixed distance band of approximately 645\,m (identified via Incremental Spatial Autocorrelation) and FDR correction, 4{,}373 ESM points formed significant hot or cold spots. As expected for a dense network, a subset of points in the inner city had very high neighbor counts (\textgreater 1000), which increases the effective spatial smoothing in these areas without invalidating local contrasts.

The following 12 locations correspond to all statistically significant hot and cold spots of collective travel experience identified by the Optimized Getis-Ord $Gi^{*}$ analysis; no locations were manually selected or excluded. 

\begin{enumerate}
\item \texttt{Spot 1} (664 reports, 131 participants), cold spot
\item \texttt{Spot 2} (82 reports, 9 participants), cold spot
\item \texttt{Spot 3} (140 reports, 28 participants), cold spot
\item \texttt{Spot 4} (1252 reports, 187 participants), cold spot
\item \texttt{Spot 5} (56 reports, 17 participants), cold spot
\item \texttt{Spot 6} (115 reports, 32 participants), cold spot
\item \texttt{Spot 7} (212 reports, 74 participants), cold spot
\item \texttt{Spot 8} (92 reports, 25 participants), hot spot
\item \texttt{Spot 9} (266 reports, 96 participants), hot spot
\item \texttt{Spot 10} (142 reports, 42 participants), hot spot
\item \texttt{Spot 11} (68 reports, 23 participants), hot spot
\item \texttt{Spot 12} (449 reports, 89 participants), hot spot

\end{enumerate}

Event patterns differ systematically between hot and cold spots (Figures~\ref{fig:eventsperspottype} and \ref{fig:hotspotsevents}). Cold spots are characterized by a concentration of negative events: delays, overcrowding, disruptive passengers, feeling unwell, and infrastructure issues are consistently higher than in hot spots. This is  further endorsed by similar free text answers, see Table \ref{tabColdfree}. By contrast, the incidence of positive events such as having arrived on schedule or time well spent is only modestly higher in hot spots and remains present even in several cold spots. This asymmetry suggests that the accumulation of negatives, rather than the simple absence of positives, is a primary driver of low collective experience ratings.

\begin{table}[H]
\centering
\renewcommand{\arraystretch}{1.05}
\begin{tabular}{p{5.5cm} p{1.2cm} p{7cm}}
\toprule
\textbf{Category} & \textbf{\%} & \textbf{Example free text} \\
\midrule
Delay/time pressure/itinerary adjustment/traffic jam & 20.0\% & ``current delay of 11 min'', ``train not coming'' \\
Bad smell or air/noisiness/crowdedness & 17.8\% & ``unpleasant smell'', ``crying children'' \\
Things went smoothly (transfer / punctuality / arrival) & 12.1\% & ``caught an earlier train'', ``got my train connection'' \\
App and study logistics & 11.3\% & ``short break for groceries'',  \\
Others & 8.2\% & ``did not get my favourite seat'' \\
In-vehicle temperature (negative) & 6.9\% & ``very warm'', ``air conditioning set too cool'' \\
Disruptive people & 4.9\% & ``3 beggars'' \\
Empty vehicle/calm surroundings/found seating/comfortable & 4.9\% & ``unusually empty'', ``it's very quiet on the bus'' \\
Wrong or too little information & 3.5\% & ``Poor signage for the side entrance'' \\
Dirty vehicle or station & 2.2\% &  ``pigeons and their droppings'' \\
Service frequency/long transfer/long waiting times & 2.2\% & ``The frequency of the S-Bahn at this time of day is highly questionable'' \\
Weather (positive) & 2.2\% & ``nice weather'', ``sun is shining'' \\
Too many stairs or long ways & 2.0\% & ``steep stairs like hiking mountains'' \\
\bottomrule
\end{tabular}
\caption{Frequency and examples of free text answer categories in all cold spots. Categories with less than 2\% were driving behavior/bus driver did not wait or stop/left too early (0.8\%), found seating (0.5\%), and weather (negative)/missing protection (0.3\%). 594 free text answers were given in total in the hot spots.}
\label{tabColdfree}
\end{table}

Spot-level inspection shows that cold spots are not homogeneous. \texttt{Spot 5} concentrates multiple issues simultaneously: overcrowding (41.1\,\%), disruptive people (33.9\,\%), infrastructure issues (25.0\,\%), and feeling unwell (25.0\,\%), with no single report of positive interaction.

On the other hand, \texttt{Spot 3} is distinguished chiefly by reliability problems (delays 26.4\,\%), indicating a delay-dominated cold spot. \texttt{Spot 6} shows comparatively high missed connections (8.7\,\%) together with disruptive people (23.5\,\%). In contrast, \texttt{Spot 4}, as the central hub in Hamburg, exhibits a broad problem profile (overcrowding 18.2\,\%, disruptive people 17.0\,\%, feeling unwell 13.7\,\%) typical of a major interchange.

\begin{table}[H]
\centering
\renewcommand{\arraystretch}{1.05}
\begin{tabular}{p{5.5cm} p{1.2cm} p{7cm}}
\toprule
\textbf{Category} & \textbf{\%} & \textbf{Example free text} \\
\midrule
App and study logistics & 25.7\% & ``currently waiting for the bus'',  \\
Bad smell or air/noisiness/crowdedness & 14.1\% & ``crowded train due to delay'' \\
Empty vehicle/calm surroundings/found seating/comfortable & 13.1\% & ``Not too crowded, not too hot, everything is relaxed today.'' \\
Delay/time pressure/itinerary adjustment/traffic jam & 7.8\% & ``closure or unserved stop'' \\
Others & 7.8\% & ``hay fever'' \\
Things went smoothly (transfer / punctuality / arrival) & 7.8\% & ``all good'' \\
In-vehicle temperature (negative) & 5.8\% & ``it's cold'' \\
Weather (positive) & 4.9\% & ``good weather'' \\
Disruptive people & 3.4\% & ``drunken beggars in the subway'' \\
driving behavior/bus driver did not wait or stop/left too early& 2.4\% & ``Bus driver did not let passenger get off despite calls'' \\
\bottomrule
\end{tabular}
\caption{Frequency and examples of free text answer categories in all hot spots. All reported categories are included regardless of valence, reflecting the full range of experiences travelers reported in these locations; the low frequency of 
negative categories relative to cold spots is itself informative. Categories with less than} 2\% were Service frequency/long transfer/long waiting times (1.9\%), Wrong or too little information (1.9\%), Dirty vehicle or station (1.5\%),polite staff (1.0\%),Too many stairs or long ways (0.5\%) and weather (negative)/missing protection (0.5\%). 206 free text answers were given in total in the hot spots.
\label{tabHotfree}
\end{table}

Hot spots exhibit distinct profiles of positive experience, see Figure \ref{fig:hotspotsevents} and Table \ref{tabHotfree}. \texttt{Spot 8} pairs very high comfort (60.9\,\%) with a pleasant environment (22.8\,\%), minimal overcrowding (1.1\,\%) and near-absence of delays and missing information, resembling a comfort-centered suburban profile. \texttt{Spot 9} combines strong time efficiency (arrived on schedule 39.8\,\%, time well spent 38.0\,\%) with comfort (44.4\,\%) and positive interactions (10.5\,\%). \texttt{Spot 10} and \texttt{Spot 11} highlight the role of urban context (nice environment 16.2\,\% and 11.8\,\%, respectively) alongside low negative events. Notably, \texttt{Spot 12} remains a hot spot despite moderate disruptive people (11.8\,\%) and overcrowding (16.0\,\%), implying that sufficiently strong positives (comfort 39.4\,\%, nice environment 12.7\,\%) can offset some negatives.

Across locations, arrived on schedule shows relatively small differences between spot types (for example, 35.5\,\% at \texttt{Spot 1} [cold] vs.\ 38.0\,\% at \texttt{Spot 8} [hot]; 37.9\,\% at \texttt{Spot 4} [cold] vs.\ 39.8\,\% at \texttt{Spot 9} [hot]). Similarly, time well spent often overlaps (for example, 33.1\,\% at \texttt{Spot 4} [cold] vs.\ 38.0\,\% at \texttt{Spot 9} [hot]). These patterns reinforce that punctuality and productive time, while important, are not sufficient discriminators. Instead, cold spots are marked by the co-occurrence of several negatives, whereas hot spots feature a bundle of positives, especially comfort, social climate (positive interactions), and a nice environment.

\newpage

\section{Discussion}

This study combined high-frequency in-situ experience sampling with spatial hot and cold spot analysis to examine how positive and negative travel experiences are distributed across an urban public transport network. The results show that travel experience is not evenly spread across the city but clusters in specific hot and cold spots, and that these locations differ markedly in the types of experiences that dominate them. The considerable within-person variance in travel experience ratings indicates that hot and cold spot patterns reflect genuine spatial variation in service quality rather than the self-selection of inherently satisfied or dissatisfied passengers into different areas. The co-occurrence of high and low ratings within hot and cold spots is unsurprising given that data were collected over seven months, spanning variation in time of day, day of week, and season. The Getis-Ord $Gi^{*}$ statistic captures a statistical tendency across these varying onditions, rather than implying that every trip in a cold spot is negative or every trip in a hot spot is positive. 

A first key finding is that cold spots are heterogeneous and problem-specific. Even though participant and report numbers vary by location, the results hint that some areas such as \texttt{Spot 3} are primarily affected by delays, while others such as \texttt{Spot 5} are dominated by overcrowding, disruptive passengers, and infrastructure issues. Major hubs such as \texttt{Spot 4} combine multiple stressors, including crowding, disruptive behavior, and feeling unwell. This suggests that cold spots do not arise from a single underlying factor but rather from different constellations of issues. Consequently, interventions should be locally tailored rather than one-size-fits-all. Measures to improve reliability are most relevant in delay-dominated cold spots, while crowd management, staff presence, or environmental design may be more effective where overcrowding and social stressors prevail.

A second important result is that also hot spots do not follow a uniform pattern but instead represent different models of positive travel experience. \texttt{Spot 8} illustrates a comfort-oriented profile with very high ratings for comfort and pleasant surroundings, combined with very low reports of negative events. \texttt{Spot 9}, in contrast, is characterized by time efficiency, with high proportions of on-time journeys and reports of time well spent, while still offering a comfortable and socially positive environment. \texttt{Spot 10} and \texttt{Spot 11} demonstrate the contribution of urban design, with relatively high reports of a nice environment and low levels of negative experiences. These examples show that positive travel experiences can emerge through different pathways, emphasizing comfort, efficiency, or environment, and each pathway may serve as a model for replication in similar contexts.

A third contribution of the study is the observed asymmetry between positive and negative events. Positive experiences such as having arrived on schedule or spending time well are common in both hot and cold spots, yet cold spots accumulate multiple negatives that appear to outweigh these positives. This suggests that worthwhile travel time, as conceptualized by \citet{cornet2022worthwhile}, can only be realized once basic functional expectations are consistently met. This asymmetry is consistent with research on negativity bias \citep{rozin2001negativity} and loss aversion \citep{novemsky2005boundaries}, which suggests that negative events exert a stronger influence on overall evaluations than positive ones. For operators, this implies that reducing a few critical negatives may lead to greater improvements in passengers' travel experience than adding new positive features. 


These findings broaden established research into transport satisfaction by presenting a fine-grained, spatially explicit perspective that complements aggregate attribute-based studies \citep{van2018influences}. Earlier work has emphasized the importance of reliability and comfort in shaping overall satisfaction \citep{redman2013quality}, but our results reveal how these factors cluster in distinct geographic patterns shaped by local infrastructure and behavior. This spatial differentiation bridges affective geovisualization approaches \citep{nold2009emotional, resch2014urban} with behavioral models of travel experience and mode choice \citep{lim2024effects}, providing a methodological foundation for real-time, passenger-centered transit planning.  

From a practical perspective, the spatial specificity of cold spots provides a triage map for targeted investigation. Locations such as \texttt{Spot 5} or \texttt{Spot 3} emerge as priorities for follow-up, and the dominant event categories reported there suggest the broad nature of the problem: reliability issues in some locations, crowding and social stressors in others. Translating these spatial signals into concrete interventions would however require more detailed investigation, for example through focus groups or triangulation with operational data on delays, occupancy, and disruptions. The primary contribution of this approach is therefore the identification of where problems consistently occur and what type of problem dominates.

Several limitations should be acknowledged. In dense parts of the city, some points had very high neighbor counts, which increases spatial smoothing and may mask micro-level variations. Event reporting relied on self-assessment and may reflect individual thresholds, making triangulation with operational data on delays, occupancy, and disruptions desirable. Furthermore, although the experience sampling approach mitigates memory bias, temporal factors, such as a tendency to respond only when the journey is going smoothly, may still introduce selection effects into the data \citep{scollon2003experience}. Future work should employ multilevel modeling to capture the nested structure of the data instead of relying on the mean value of the 8 travel experience questionnaire items \citep{muthen2023using}. 

In conclusion, the combination of real-time experience sampling and spatial statistics reveals concentrated clusters of both negative and positive passenger experiences. Cold spots arise from different constellations of problems and call for locally specific solutions, while hot spots demonstrate multiple pathways to high satisfaction that can inform best practices. Improving public transport by removing key negatives and reinforcing demonstrated positives such as comfort, social climate, and environmental quality is essential for making sustainable modes of travel an attractive choice.

\section{Declaration of generative AI and AI-assisted technologies in the writing process.}

Statement: During the preparation of this work, the authors used ChatGPT 5 (Version September 2025) and Claude Sonnet 4.6 (Version April 2026) to support spell-checking and refine wording. After using this tool, the authors reviewed and edited the content as needed and take full responsibility for the content of the published article.

\section{Declaration of Interest.}
The authors declare no conflict of interest.

\section{Acknowledgements.}
The authors thank Dominik Radzuweit for enabling the data collection by providing relevant contacts and contextual information. We also thank Jakob Dietze for his support with the data acquisition, and Jakob Dietze and Lene Manzau with categorization of the free-text answers.

CRediT roles: Conceptualization, Data curation, Formal analysis: EB; Funding acquisition: KI; Investigation, Methodology, Project administration, Validation: EB; Software: ASB; Visualization: MS and EB; Writing – original draft: EB; Writing – review and editing: EB, MS and KI.

Funding: This work was supported by mFUND of the German Federal Ministry of Transportation under the project 'Erlebensatlas' \footnote{https://www.dlr.de/en/ts/research-transfer/projects/erlebensatlas}, grant number 19F1197A.

\newpage

\bibliographystyle{elsarticle-harv} 
\bibliography{02_references.bib}

\end{document}